\documentclass[aps,prb,twocolumn,groupedaddress]{revtex4}
\usepackage{graphicx}
\usepackage{psfrag}
\usepackage{bm}
\usepackage{dcolumn}

\begin{document}

\title{Edge capacitance of a 2D topological insulator. }

\author{M.V. Entin}
\affiliation{Rzhanov Institute of Semiconductor Physics, Siberian
Branch of the Russian Academy of Sciences, Novosibirsk, 630090, Russia}
\affiliation{Novosibirsk State University, Novosibirsk, 630090, Russia}
\email{entin@isp.nsc.ru}

\author{L. Braginsky}
\affiliation{Rzhanov Institute of Semiconductor Physics, Siberian
Branch of the Russian Academy of Sciences, Novosibirsk, 630090, Russia}
\affiliation{Novosibirsk State University, Novosibirsk, 630090, Russia}
\email{brag@isp.nsc.ru}

\begin{abstract}
We study capacitance of the 2D topological insulator (TI) edge states. The total capacitance is combined as a serial circuite of 3 capacitors presenting geometrical $C_G$, quantum $C_Q$ and correlation $C_{corr}$ contributions to the electron energy. If the  Coulomb interaction is weak, they obey an inequality  $C_G<C_Q<C_{corr}$. Quantities $C_G$ and $C_Q$ are found in the case of a round TI dot.  The quantum capacitance at the finite temperature is determined taking into account the  edge states quantization  with and without the magnetic field. We have concluded that, in the accepted approximations, $C_{corr}=0$.
\end{abstract}

\maketitle

\subsection*{Introduction}
The 2D topological insulator (TI) is one of the most actively developing area of  solid state physics (see, e.g., reviews \cite{bern}-\cite{konig} and the references therein). One of the most known  2D TI is the 2D CdTe/HgTe/CdTe quantum well with  HgTe layer width $d>$ 6.3 nm.
  The  electron spectrum of the unbound 2D TI has an energy gap, while the spectrum of the bound one is gapless  due to inevitable existence  of the  edge states in  the gap.  It is important that dispersion of the edge states is linear in the wide range of the electron momentum near zero.

Capacitance measurements is a premier
tool to study the electronic properties of 2D systems. This method was applied also to 2D TI
 \cite{Kernreiter2016}, \cite{Kozlov2016}.  In particular, the capacitance determines the Coulomb blockade energy when the transport through the quantum dots is studied. Hence, besides the direct measurements of the  capacitance, it can be measured indirectly via the Coulomb blockade. In the ideal topological insulator the edge states are the only in the energy gap that recharges when the voltage is applied between the gate and  TI layer. This process determines the capacitance of the 2D topological insulator that is the subject of this paper.  We study the systems with infinite edges like a 2D strip or close edge like a round TI spot surrounded by a usual insulator where the edge states are formed on the TI circumference.

We  consider 3 contributions to the capacitance. First, it is  geometrical capacitance $C_G$. Roughly, the system can be treated as a 1D metallic wire with some length $L$ and width $1/\kappa$ situated at the distance $d$ near the metallic gate or a plate capacitor with  area $ L/\kappa$. The first model is valid, if $h\kappa\gg 1$, and gives $1/C_G\sim L \ln (\kappa h)/\epsilon $  (here $\epsilon$ is the dielectric constant). The second one gives  $C_G=\epsilon L/(4\pi h\kappa  )$ and is valid at $h\kappa\ll 1$.  The term ''geometrical capacitance'' shows that it does not include any quantum sizes, except the edge state width.

The other part of the capacitance is a ''quantum'' one $C_Q$ \cite{luryi}. This capacitance originates from the Fermi energy of free electron gas and at low temperature universally depends on the density of states (DOS) $\varrho$: $C_Q=L\varrho/ e^2 $. In the quantum wire with linear spectrum $C_Q= const$.

The third part is the correlation capacitance $C_{corr}$ that originates from the electron interaction. In the bulk of a metal the correlation energy is of the order of $e^2/r_d$ ($r_d$ is the Debye radii), so that $ C_{corr}\propto 1/e^3$. Thus, $C_{corr}\gg C_Q\gg C_G$ if the e-e interaction is weak ($E_F \gg E_B$). For the edge states, however, the correlation energy should be calculated separately. Although the interaction weakness is necessary for the application of the perturbation theory, actually, the perturbation parameter is not so small.  Therefore, the additional terms have to be taken into account. In this research we consider the terms up to the order of the correlation energy.

The electron energy is additive.  Hence, the inverse capacitances are additive too, and all 3 capacitances compose a series circuit.

In this paper we  determined the geometric capacitance of a round TI in a gated 2D TI and found this value to be proportional to the TI length in the limit of a large radius.   We found the quantum capacitance of the round 2D TI  taking into account the longitudinal  edge states quantization, finite temperature, and  normal magnetic field. The correlation energy and correlation capacitance of the straight edge electron liquid were   found to be vanished in our approximation.
\subsection*{Problem formulation}
We deal with an ideal 2D topological insulator, (HgTe) embedded into the usual semiconductor (CdS),  covered by a metallic gate at the distance $h$ from the HgTe layer, supposed to be the $XY$ plane. The Fermi level is situated in the gap of the 2D TI. In this situation, the gate controls the  charging of the edge states only. We use the  Bernevig-Hughes-Zhang model of 2D TI \cite{bern}. Dispersion of electrons in the straight-edge states is  $\varepsilon_\sigma(p)=E_0+\sigma v p$, where $\sigma=\pm$, $p$ is the 1D momentum along the edge $y=0$, $E_0$ is the G-point energy, and $v$ is the band velocity. (The units where $\hbar=k_B=1$ are accepted.) In the case of the straight edge the TI is situated at $y>0$.  The envelope wave function of the electrons at the edge state is $e^{ipx}\chi(y)\xi(z)$, where $\chi(y)=Z(e^{-\lambda_1y}-e^{-\lambda_2y})$,
\[
Z=\sqrt{\frac{2 \lambda_1\lambda_2(\lambda_1+\lambda_2)}{(\lambda_1-\lambda_2)^2}},
\]
 and $\xi(z)$ is some localized wave function of the electron motion across the layer. For  the HgTe quantum well of the 7 nm width, the 2D TI band parameters are $\lambda_1\approx A/\sqrt{B^2-D^2}$, $\lambda_2\approx M\sqrt{B^2-D^2}/(AB)$, where  $A = 364.5$meV$\cdot$ nm,  $B = 686$meV$\cdot$ nm${}^2$, $D = 512$meV$\cdot$ nm${}^2$, $M = 10$meV \cite{konig}.
Thus, $\lambda_1/ \lambda_2=41.7\gg 1$; this allows  neglecting $e^{-\lambda_1 y}$ in the wave function $\chi(y)$.

As far as the geometric capacitance of the infinite wire diverges in the absence of the metal gate, we  consider a circular edge. The results are applicable at the edge of a round TI quantum dot surrounded by  an usual insulator. They can be also applied for the direct edge, if  $R\gg h$ and $R\gg 1/\lambda_2$, where $R$ is the  TI radii.

We found the geometric capacitance  assuming the electrons to be uniformly distributed along the edge.
The quantum capacitance is determined by the variation  of  edge charge $eN$ with the variation of the Fermi level $\mu$: $e\delta N/\delta \mu$.
The correlation capacitance mainly depends on the local redistribution of electrons along the edge state. It is found in the model of the straight edge.

\subsection*{Geometric capacitance}

Considering a round TI, we assume $y=R-r\sim 1/\lambda_2$ and the charge density  $\rho(r)=\rho_0\exp[2\lambda_2(r-R)]\theta(R-r)$, where $(r,\varphi)$ are the polar coordinates.
The self-energy of this distribution is
\begin{equation}\label{100}
  W=\frac{1}{2\epsilon}\int dq(r)\,dq(r') \left(\frac{1}{|{\bm r}-{\bm r}'|}-\frac{1}{|{\bm r}-{\bm r}'-2{\bm h}|}\right).
\end{equation}
Here $dq(r)=\rho(r)r\, dr$ and ${\bm h}=(0,0,h)$, $h$ is the distance  to the gate. The second term in the integrand is due to mirror reflection of the ring in the gate electrode.
Assuming $h\ll R$ and $\lambda_2R\gg 1$,  we obtain
\begin{equation}\label{400}
W=\frac{\sqrt{\pi}R\rho_0^2}{4\epsilon\lambda_2^2}G_{01}^{32}\left(4\lambda_2^2h^2\left|^{1,1}_{\frac{1}{2},1,1,0}\right.\right)
\end{equation}
Here $G_{01}^{32}\left(4\lambda_2^2h^2\left|^{1,1}_{\frac{1}{2},1,1,0}\right.\right)$ is the Meijer function. Therefore,
\begin{equation}\label{350}
  C_G=\frac{2\pi^{3/2}\epsilon R}{G_{01}^{32}\left(4\lambda_2^2h^2\left|^{1,1}_{\frac{1}{2},1,1,0}\right.\right)}
\end{equation}

Using the expansions of the Meijer function for $\lambda_2 h \ll 1$ and $\lambda_2 h \gg 1$, we find

\begin{equation}\label{450}
   C_G=\left\{\begin{array}{cc}
              \frac{\epsilon R}{2\lambda_2 h}, & \mbox{  if  }   \lambda_2 h \ll 1 \\
              \frac{2\pi\epsilon R}{\log( 4\lambda_2 h)+\gamma}, & \mbox{  if  }   \lambda_2 h \gg 1
            \end{array}
   \right.
\end{equation}
here $\gamma$ is the Euler constant.

Note that $C_G$ in Eqs.~(\ref{350}, \ref{450}) is proportional to the ring length $2\pi R$, so that this value seems to be independent of the TI outline. Actually, this dependence arises from the terms of the order of $\lambda_2R$ and $h/R$ that were omitted in Eqs.~(\ref{400}, \ref{350}).

\subsection*{Quantum capacitance}

The quantum capacitance is described by the S. Luriy formula \cite{luryi}
$ C_Q=e^2\varrho$,
where $\varrho$ is the DOS of the edge. For the edge of length $2\pi R$ $\varrho=2R/\hbar v$ and
\begin{equation}\label{quant}
    C_Q=2e^2R/\hbar v.
\end{equation}
The last expression is valid if the edge states quantization is negligible. For example, if the system contains many such edge-states-rings of   varying sizes, this formula yields the mean capacitance per  ring. The same is true for a single ring if its temperature exceeds the distance between the edge states levels.

 Let us consider the quantum case when the distance between levels is essential. We   include also a normal magnetic field $B$. The edge states compose a set
 \begin{equation}\label{En}
   E_{n,\sigma}= n\omega+\sigma \omega_B/2, ~~~~\sigma=\pm 1.
\end{equation}
Here $n$ is an integer, $\omega=v/R$ is the distance between the same-spin edge states, $\omega_B=veBR$ characterizes the splitting of contra-propagating edge states by the magnetic field. The characteristic value of $\hbar\omega=v/R$ at $v=3.7 \cdot 10^7$cm/s and $R=1$ $\mu$m is $\omega=5.3$K.

The DOS reads
 \begin{equation}\label{quantDOS}
  \varrho(\mu)=\sum_{n,\sigma}\delta(\mu-E_{n,\sigma})
\end{equation}
To include the levels width $\gamma$, one should replace the $\delta$-function with the Lorenzian $\delta(E)=\gamma/\pi(E^2+\gamma^2).$
At a finite temperature the quantum capacitance has to be found taking into account the Fermi distribution. Then, instead of Eq. \ref{quantDOS} we obtain

\begin{equation}\label{quantDOST}
  \varrho(\mu)=\frac{1}{T}\sum_{n,\sigma}\frac{e^{(E_{n,\sigma}-\mu)/T}}{(e^{(E_{n,\sigma}-\mu)/T}+1)^2}
\end{equation}
If $\mu$ coincides with a level, the total capacitance approaches the geometrical one. If $\mu$ is situated  between the levels, it approaches the quantum capacitance.
At high temperature $T$ (\ref{quantDOST}) gives (\ref{quant}) (independently of the magnetic field). The quantum corrections to $\varrho(\mu)$  can be also found by the Fourier expansion. As a result, the corrected formula yields
\begin{eqnarray}\label{quant-corr}
    &&C_Q=\frac{2e^2}{\hbar v} R\left[1+\sum_{n\neq 0,\sigma}\cos\left(2\pi n\frac{\mu+\sigma\omega_B/2}{\omega}\right)\right.\\
    &&\left. \times\tilde{f}\left(\frac{2\pi n T}{\omega}\right)\right],
    \;\;\tilde{f}(p)=\int dx e^{ipx}\frac{e^x}{(e^{x}+1)^2}=\frac{\pi p}{\sinh(\pi p)}\nonumber
   \end{eqnarray}

The capacitance oscillates with the Fermi level $\mu$ and  magnetic field $B$.  The oscillations exponentially decay at high temperature ($T>1$K) and only the  term with $n=1$ remains in the oscillating contribution.
At $T\gg 1$K the oscillating term in $C_Q$ is small as compared with the regular one.

 Expanding $C_Q$ in series of $B$ we find a weak-field magnetocapacitance
     $$C_Q^B=-\frac{4\pi^2e^2}{\hbar v} R\frac{\omega_B^2}{\omega^2}\cos(2\pi\mu/\omega)\frac{2\pi  T}{\omega \sinh(2\pi T/\omega)}.$$
The weak-field magnetocapacitance oscillates only  as a function of $\mu$.

Here we have considered  the round quantum dot of TI. Actually, the results do not essentially  depend on the  dot shape, if  $\omega\to v/L$ and $\omega_B=eBS/c$, where $L$ and $S$ are the quantum dot perimeter and area, respectively. These expressions are valid, if $L\gg 2\pi\hbar/p_F$. Here  $p_F$ is the Fermi momentum, so that  $2\pi\hbar/p_F$ is the electron wavelength.

\subsection*{ Correlation energy }

   The energy of interaction of two electrons on a straight edge  is

\begin{eqnarray}\nonumber&&V(x)=\frac{e^2Z^4}{\epsilon}\int_0^\infty
dydy_1dzdz_1 |\chi(y)\xi(z)\chi(y_1)\xi(z_1)|^2
\\&&\times\left(\frac{1}{r}-\frac{1}{|{\bf r}-{\bf
h}|}\right).\label{V}\end{eqnarray}
Here $x$ is the distance between the electrons, ${\bf
r}=(x,y-y_1,z-z_1)$, ${\bf h}=(0,0,h)$.

This  interaction is diagonal with respect to  spin index  $\sigma,\
\sigma'$. Therefore, for the total interaction, we can write
$V_{\sigma\sigma'}(x)= \delta_{\sigma,\sigma'}V(x)$.

The Fourier transform of  1D potential (\ref{V}) is
\begin{eqnarray}\nonumber V(q)=\frac{4\pi e^2Z^4}{\epsilon}\int
\frac{dk_ydk_z}{4\pi^2}\frac{(1-e^{-2hik_z})}{k^2}|\eta(k_z)|^2|\zeta(k_y)|^2,
\end{eqnarray}
where   $k^2=q^2+k_y^2+k_z^2$,
$$\eta(k_z)=\int\frac{dz}{2\pi}|\xi(z)|^2e^{-ik_zz},~~~\zeta(k_y)=\int\frac{dy}{2\pi}|\chi(y)|^2e^{-ik_yy},$$
$$\zeta(k_y)=\frac{4(\lambda_1-\lambda_2)^4}{(4\lambda_1^2+k_y^2)(4\lambda_2^2+k_y^2)((\lambda_1+\lambda_2)^2+k_y^2)}.$$

Asymptotics of this equation are
\[
V(q)=\left\{\begin{array}{cc}
              \frac{\pi e^2Z^4\kappa h}{\epsilon}, & h\to 0, \\
              \frac{2 e^2\kappa}{\epsilon q} \arccos
   \left(\frac{\kappa}{\sqrt{q
    ^2+\kappa^2}}\right), & h\to \infty.
            \end{array}
\right.
\]

The equations above represent a bare Coulomb interaction and do not take
into account the edge-state electron polarization.

    The capacitance is determined by the $\Omega$ -potential of the
interacting edge 1D electron gas $$C=e^2\frac{\partial^2
\Omega}{\partial \mu^2}$$
    The basic interaction corrections to the $\Omega$ -potential of the
edge 1D electron gas are given by
    $$\Delta\Omega_1=-\frac{L}{2\pi^2}\int^{p_F}\int^{p_F}V(p-p')dpdp'$$
    $$C=-\frac{e^2L}{2v^2\pi^2}V(q=0)$$
This is just the geometric capacitance Eq. (\ref{350},\ref{400}).

To obtain the corrections due to correlation,  we need to calculate the
subsequent orders over the interaction constant. The nearest order is
determined by the renormalized Coulomb interaction.
It is given by the polarization operator of 1D electron gas
$\Pi(q,\omega)$. For the 1D system with the linear spectrum, we obtain
\begin{eqnarray}\label{pol}
&&\Pi_\sigma(q,\omega)=2\delta_{\sigma\sigma'}\int\frac{dp}{2\pi}\frac{f(p-q)-f(p)}{\varepsilon_\sigma(p-q)-\varepsilon_\sigma(p)+\omega}\\
 &&   =\frac{q}{\pi(\omega-\sigma vq)}, \nonumber
\end{eqnarray}
where $f(p)$ is the Fermi function.
 The vertex part
is
$$\Gamma_{\sigma\sigma'}(q,\omega)\equiv\Gamma_{\sigma}(q,\omega)\delta_{\sigma\sigma'},~~~\Gamma_{\sigma}(q,\omega)=\frac{V(q)}{1-\Pi_\sigma(q,\omega)V(q)}.$$
It replaces $V(q)$ in the next order of the perturbation.

The relevant correlation to the $\Omega$-potential at a low temperature  per
unit length is \cite{Abrikosov}
\begin{eqnarray}\label{correl}
   \Omega_{corr}
=-\frac{L}{2\pi}\sum_\sigma\int_0^{e^2}d(e^2)\nonumber\\\times\int_{-\infty}^{\infty}
d\omega dq
e^{i\omega\tau}\Pi_\sigma^2(q,i\omega)\Gamma_\sigma(q,i\omega)|_{\tau\to+0}.
\end{eqnarray}
Calculating the integrals over $\omega$ in Eq.(\ref{correl}) assuming
$V(q)>0$,  we find $\sum_\sigma\int d\omega\propto
\theta(V(q)-v)/V(q)q$. Therefore, $\Omega_{corr}=0$ at small
$e
V(q)<v$,  at least, in the considered order. Thus, we can set
$\Omega_{corr} =0$ and $C_{corr}=\infty$.

Hence, electrons do not redistribute by an interaction
potential. This results from the linearity of the electron spectrum: in such case in the framework of the  Schr\"odinger equation the electron density  is not affected by an external potential.
The details are the subject of the next publication.

 \subsection*{Conclusions}
Here we have determined  the inverse edge capacitance of the topological insulator $1/C=1/C_G+1/C_Q+1/C_{corr}$ composed from 3 contributions: geometrical $C_G$ [Eq.~(\ref{350})], quantum $C_Q$ [Eq.~(\ref{quant-corr})], and correlation $C_{corr}$ capacitances.
For a weakly interacting electron system, the main contribution is the geometrical capacitance. Roughly, $C_G$ is  proportional to the length of the edge divided by $\log( 4\lambda_2 h)$ [Eq.~(\ref{450})].
The quantum capacitance [Eq.~(\ref{quant-corr})] is usually small due to the interaction constant $g=e^2/\kappa\hbar  v$, but it does not contain the mentioned logarithm. Hence, the quantum contribution could be comparable with the geometrical one, if $g\log( 4\lambda_2 h)\simeq 1$ (this condition hardly can be satisfied).  The last contribution $1/C_{corr}$ caused by the interaction between electrons  is vanishing in our approximation.   It should be emphasized that the main contribution exceeds the quantum contribution only logarithmically. This differs the result from the case of the large 2D quantum dot, where this factor is proportional to the quantum dot area. That makes the Coulomb energy of a large quantum dot essentially exceeding the level spacing; the source of this difference is the 1D nature of the edge state.

We consider the quantum capacitance of a confined edge state taking into account its quantization. In particular, we have determined its oscillatory behavior at a finite temperature, the magnetocapacitance  at a low magnetic field, and  magnetic oscillations of capacitance at a finite magnetic field.

\begin{figure}
\begin{center}
\includegraphics{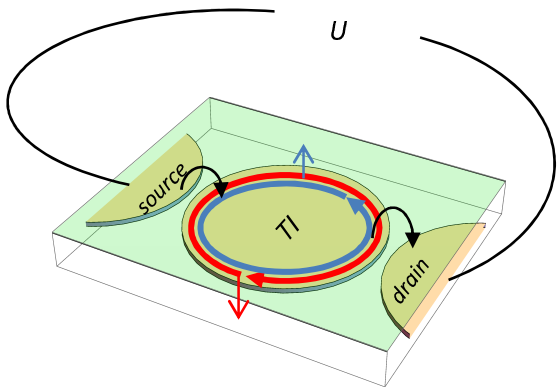}
 \end{center}
\caption{ \label{Fig.Model1} (Color online) The sketch of the Coulomb blockade via a round TI dot connected to two contacts. All area is
covered with the gate controlling the Fermi
level position. Electrons  go from the source to drain  through the edge states. }
\end{figure}

Discuss now the consequences of the edge state capacitance for the Coulomb blockade.
The edge capacitance  affects the electron transport in a structured 2D TI, e.g., in  a system with a 2D TI quantum dot (see Fig 1.), where the edge capacitance gives the Coulomb blockade energy  for an electron hopping via the quantum dot $e^2/2C$. Similar barriers appear in the  uniform 2D TI with a fluctuating gap sign. In such  system at a low temperature electrons jump between the edges, and the random Coulomb blockade energy should determine the Coulomb-gap-mediated hopping conductivity. This problem goes beyond the scope of this paper and will be studied  later.
\subsection*{Acknowledgements}
This research was supported  by RFBR grant  No 14-02-00593.

 \end{document}